\documentclass[a4paper,14pt]{extarticle}

\usepackage{graphicx}
\usepackage{amsmath,amsthm,amssymb}

\usepackage[bindingoffset=1cm,
            left=1.4cm,
            right=1.4cm,
            top=1.4cm,
            bottom=2cm,
            footskip=1cm]{geometry}
\usepackage{float}
\usepackage{enumitem}
\usepackage{multirow}

\newcommand{\beq}{\begin{eqnarray}}
\newcommand{\eeq}{\end{eqnarray}}
\newcommand{\be}{\begin{equation}}
\newcommand{\ee}{\end{equation}}

 \def\la{\mathrel{\mathpalette\fun <}}

\def\fun#1#2{\lower3.6pt\vbox{\baselineskip0pt\lineskip.9pt
\ialign{$\mathsurround=0pt#1\hfil ##\hfil$\crcr#2\crcr\sim\crcr}}}

\newcommand{{\SD}}{\rm SD}

\newcommand{{\Mc}}{\mathcal{M}}

\newcommand{\vep}{\mbox{\boldmath${\rm p}$}}

\newcommand{\ves}{\mbox{\boldmath${\rm s}$}}

\newcommand{\lan}{\langle}
\newcommand{\ran}{\rangle}

\title{The Spin-Spin Dynamics of Glueballs\thanks{v1 (for arXiv) -- \today}}
\author{{A.M.~Badalian}\thanks{badalian@itep.ru},\quad {M.S.~Lukashov}\thanks{m.s.lukashov@gmail.com} \bigskip \\ NRC ``Kurchatov Institute'' \\ Moscow 123182, Russia}
\date{\today}

\begin{document}
\maketitle

\begin{abstract}
The masses of pure gauge glueballs are calculated with the use of relativistic string Hamiltonian without fitting parameters. The string tension $\sigma_f=0.184$~GeV$^2$ in fundamental representation is fixed, using the Necco-Sommer lattice data, and to calculate the vector coupling $\alpha_{\rm V}(r)$ the value of $\Lambda_{\overline{MS}}^0=238$~MeV ($N_f=0$) is taken. The spin-spin potential, defined via the vacuum correlation function, is shown to produce a screening effect and decrease  a hyperfine splittings between  tensor and scalar glueballs. The masses of first and second $0^{++}$, $2^{++}$ excitations are predicted. For the ground states the masses $M(0^{++})=1508$~MeV, $M(2^{++})=2292$~MeV (variant A), in agreement with those of $f_0(1500),~f_2(2300)$ are obtained, and the first excitation mass $M(0^{++})=2613$~MeV is predicted. In case B  $M(0^{++})=1.669$~MeV, $M(2^{++})=2212$~MeV are obtained.
\end{abstract}

\section{Introduction}

The glueball masses and decays were studied for decades, nevertheless,  even the lowest scalar glueball is not still identified.
In lattice QCD its mass is predicted with uncertainty $\sim 200$~MeV,  $M(0^{++})\cong (1.6-1.8)$~GeV \cite{1}-\cite{4}, and in the range (1.4--1.9) GeV in
different theoretical models \cite{5}-\cite{11}. Recently the BESIII has observed new resonance,
$X(2370)~(J^{PC}=0^{-+})$, in the $J/\psi \rightarrow \gamma X(2370)$ decays \cite{12}, which is supposed to be the pseudoscalar
glueball. This gives new information which may improve our understanding of the glueball spectrum as a whole. In
our paper we study first and second excitations of scalar and tensor glueballs and   concentrate on their hyperfine splittings
$\delta_{hf}$, which often contain rather arbitrary parameters. For that reason predicted $\delta_{hf}$ between ground states vary
from 440 MeV in \cite{10} up to the value 820~MeV in \cite{7}. Such large differences of $\delta_{hf}$ is characteristic feature
of glueball spectrum and absent in mesons.

In our analysis we use the relativistic string Hamiltonian (RSH)  \cite{13,14,15,16}, which was derived from the gauge-invariant Green's function with spinless
quarks (gluons) and already applied to glueballs \cite{5,6}, but here we escape any approximations, because relativistic effects are large and very important
for glueballs. In pure gauge theory ($N_f=0$) the spin-averaged glueball mass is determined by single parameter $\sigma_a$ - the string tension in adjoint representation,
which is expressed via the string tension $\sigma_f$ in fundamental representation (FR). Therefore to fix
$\sigma_f$ we use the lattice data, referring to static forth $F(r)$: the Sommer relation \cite{17} and two-loop vector coupling $\alpha_{\rm F}(r)$,
defined by the QCD vector constant, expressed via known $\Lambda_{\overline{MS}}^0=238(20)$~MeV~ ($N_f=0)$ \cite{18,19}.
Here we perform detailed analysis of the static force (in FR) and compare our results with lattice data from the Necco, Sommer paper \cite{20}.
As in \cite{20}, we take the Sommer scale $r_0=0.50$~fm and the product $r_0 \Lambda_{\overline{MS}}^0 = 0.602$. Note that last number is smaller than
$r_0\Lambda_{\overline{MS}}^0\sim 0.660(20)$~($r_0\cong 0.48$~fm), used in some other lattice analysis \cite{19}.
With $\Lambda_{\overline{MS}}^0 = 238$~MeV and the vector coupling, $\Lambda_{\rm V}^0(2-loop)=1.5995~\Lambda_{\overline{MS}}^0=380$~MeV,
we show that dependence of the coupling on $r$ is very important, and in $F(r)$ the derivative in
$\alpha_{\rm F}(r)=\alpha_{\rm V}(r) - r \alpha'_{\rm V}(r)$, is not small up to distances $\sim 0.30$~fm and
has to be taken into account.

\section{The RSH and the Static Force}

We apply the RSH \cite{14,15}, discussed in details in \cite{5,6}, to study scalar and tensor glueballs with the gluon mass $m_g=0$. As in the case of  light mesons,
the spin-spin potential is considered as a perturbation and then the RSH reduces to simple form,
\be
H_0 = 2\sqrt{\vep^2} + V_a(r), \quad\text{with }V_a(r) =\sigma_a r. ~~
\label{eq.01}
\ee
and the spin-averaged mass $M_{cog}(nL)$ is defined by the equation,
\be
H_0 \psi(r) = M_{cog}(nL) \psi(r),
\label{eq.02}
\ee
The values of $M_{cog}$ are very important for further analysis and defined via the string tension in adjoint representation,
$\sigma_a = \frac{9}{4} \sigma_f$, i.e. chosen $\sigma_f $ in FR defines the glueball mass $M_{cog}(nL)$. To fix $\sigma_f$ we use the lattice
data from \cite{20}. In mesons the value of $\sigma_f$ is usually fixed by the slope of the leading Regge trajectory, equal to $2\pi\sigma_f=1.12(3)$~GeV$^2$,
or $\sigma_f= 0.180(2)$~GeV$^2$. Below the value $\sigma_f=0.184$~GeV$^2$ ($\sigma_a=0.414$~GeV$^2$) is extracted from
analysis of the lattice data. Note that in lattice QCD (LQCD) different values of $\sigma_f$ are used, e.g. (in GeV$^2$) $\sigma_f=0.1936$ ($\sigma_a=0.436$) 
in \cite{1}, or $\sigma_f=0.234$ ($\sigma_a=0.527$) in \cite{4}.

As in \cite{20}, we calculate the force $F_f(r)=\frac{dV_f}{dr}$ and the function  $\Phi(r,r_0)= r_0^2 F_f(r)$ in the range, $(0.10~\text{fm} \leq r \leq 0.65~\text{fm})$,
staring from the static potential in FR,
\be
V_f(r) = \sigma_f r - \frac{4\alpha_{\rm V}(r)}{3 r}, ~~
\label{eq.03}
\ee
where the strong coupling in coordinate space is defined via the coupling in the momentum space \cite{21}
\be
 \alpha_{\rm V}(r) = \frac{2}{\pi} \int^\infty_0 {\rm d}q \alpha_{\rm V}(q) \frac{\sin(qr)}{q}.  ~~
\label{eq.04}
\ee
We take $\alpha_{\rm V}(q)$ in two-loop  approximation with $\Lambda_{\rm V}^0=380$~MeV (or
$\Lambda_{\overline{MS}}(N_f=0) = 238$~MeV), as in \cite{19,20,21}. Also in the coupling we use the infrared regulator
$M_B=1.0$~GeV, introduced in background perturbation theory \cite{22} to provide freezing of the coupling at large distances.

As seen from Eq.~\ref{eq.03}, the static force includes the derivative $\alpha'_{\rm V}(r)$,
\be
F_f(r) = \sigma_f + \frac{4 (\alpha_{\rm V}(r) - r \alpha'_{\rm V}(r))}{3 r^2},~~
\label{eq.05}
\ee
which is not small up to the distances $r\cong 0.60$~fm. In the force the effective coupling,
$\alpha_{\rm F}(r) = \alpha_{\rm V}(r) - r \alpha'_{\rm V}(r)$, is denoted as $\alpha_{\rm F}(r)$, where the term with the derivative
can reach 30\% at small $r$ (see Table~\ref{tab.01}).

In two-loop coupling $\alpha_{\rm V}$ the vector QCD constant is taken as: $$\Lambda_{\rm V}^0(N_f=0) = 1.5995 \Lambda_{\overline{MS}}^0 = 380 \text{ MeV.}$$
As in \cite{20}, we calculate the function $\Phi(r,r_0) = r_0^2 F_f(r) = r_0^2 \sigma_f + \frac{4}{3}\frac{r_0^2}{r^2} \alpha_{\rm F}(r)$ with $r_0 = 0.50$~fm (2.5339~GeV$^{-1}$)  and
$\Lambda_{\rm V}^0=380$~MeV. At the point $r_0$ the Sommer relation:
\be
\Phi(r_0,r_0) = r_0^2 F(r_0) = 1.65.~~
\label{eq.06}
\ee
is satisfied. Comparison our numbers and the values of $\Phi(r,r_0)$  from the Necco, Sommer paper \cite{20} shows that
good agreement (better 2\%) is reached at all distances with
\be
\sigma_f=0.184 \text{ GeV}^2.  ~~
\label{eq.07}
\ee
Note that the extracted value of $\sigma_f=0.184$~GeV$^2$ agrees with $\sigma_f$ in light mesons and gives
\be
\sigma_a=\frac{9}{4}\sigma_f= 0.414~GeV^2.~~
\label{eq.08}
\ee

\begin{table}
\caption{Calculated $\alpha_{\rm V}(r),~\alpha_{\rm F}$ and the function $\Phi(r,r_0)$ from Necco, Sommer paper \cite{20} with  $r_0=0.50$~fm}
\label{tab.01}
\begin{center}
\begin{tabular}{|c|c|c|c|c|}
\hline
\multirow{2}{*}{ $r$ (in fm)}& \multirow{2}{*}{$\alpha_{\rm V}(r)$} & \multirow{2}{*}{$\alpha_{\rm F}(r)$}& \multirow{2}{*}{$\Phi(r,r_0)$ \cite{20}}& $\Phi(r,r_0)$ \\
 & & & & {[this work]} \\
\hline \hline
 0.103         & 0.272    &  0.188   &  7.30(11)        & 7.10       \\
 0.128         & 0.294  &   0.206   &  5.363(78)        &  5.372       \\
 0.154        & 0.313  &    0.224   &   4.244(53)          &  4.305   \\
 0.180         &   0.324  &  0.237 &  3.538(48)        &  3.620    \\
 0.205         & 0.335    & 0.249   & 3.060(38)      & 3.103    \\
 0.231        & 0.341   &  0.258    &     2.713(30)   &  2.792   \\
 0.250      & 0.351    & 0.268   &  2.518(16)          & 2.600  \\
 0.30        & 0.367   & 0.292   &  2.173(15)        & 2.260   \\

 0.350      & 0.377   & 0.310    &   1.951 (14)       & 2.024     \\
 0.400       & 0.386   & 0.324      & 1.812(11)      & 1.850   \\
 0.450       & 0.391   & 0.336    &  1.722(11)     &1.734  \\
 0.50        &  0.398   & 0.352   & 1.65         &  1.65  \\
 0.550       &  0.402  & 0.361  & 1.592(10)    & 1.579    \\
 0.60         & 0.406   &0.369   & 1.559(13)   &  1.524 \\
 0.65         & 0.409   &  0.376&   1.537(18)     &  1.523 \\
\hline
\end{tabular}
\end{center}
\end{table}
We give several useful values: $\alpha_F(r=r_0)=0.352$ at $r_0=0.50$~fm, $\alpha_{\rm F}(r^*)= 0.292$ at
$r^*=0.30$~fm; also $\sqrt{\sigma_f r_0^2} = 1.10$ and $r_0\Lambda_{\overline{MS}}^0 = 0.603$. Note that in
\cite{23,24}  larger $r_0\Lambda_{\overline{MS}}^0=0.632(20)$ was used with the same $r_0=0.50$~fm; it gives
$\Lambda_{\overline{MS}}^0=249(8)$~MeV , $\Lambda_{\rm V}^0=398(13)$~MeV, and larger values of the coupling
$\alpha_F(r)$ and $\Phi(r,r_0)$. These difference can affect the values of spin-spin splitting.

Here we have not taken into account the gluon-exchanged potential $V_{ge}(r)$, since this potential is not well defined.
Namely, it is not clear whether this potential has screening factor or not and its contribution to the glueball mass
may vary from $\sim 350$~MeV (no screening) to $\sim 80$~MeV (strong screening) (our estimates).
Calculations with $V_{ge}$ were performed in \cite{5,7,8}. Here we assume that $V_{ge}$ potential is suppressed.

\section{The spin-averaged masses of scalar and tensor glueballs with $n_r=0,1,2$}

First of all, with the static potential  $V_a(r)= \sigma_a r,~(\sigma_a=0.414$~GeV$^2$)  we calculate
the spin-averaged masses $M_{cog}(nL)$ - the solutions of the Eq.~\ref{eq.02} (see Table~\ref{tab.02}) and
the masses of the $0^{++},2^{++}$ ground and excited states will be presented in next Section.  Note that
in RSH many formulas include the glueball kinetic energy $\mu(nL)$,
\be
 \mu(nL) = \frac{1}{4} M_{cog}(nL) = \frac{1}{2} (\sigma_a \lan r \ran_{nL}).  ~~
\label{eq.09}
\ee
In Table~\ref{tab.02} the masses $M_{cog}(nL)$ and the matrix elements (m.e.) $\lan r \ran_{nL}$, $\sqrt{\lan r^2 \ran_{nL}}$, $(n=n_r+1)$ are
given, where the r.m.s. of the ground state ($L=0$), is equal to 0.547 fm, i.e. significantly
smaller than that of $\rho$-meson. Also the r.m.s. of excited glueballs appear to be not large, $\la 1.0$~fm.
\begin{table}
\caption{The spin-averaged masses $M_{cog}(nL)$ (in GeV) of glueballs with $n_r=0,1,2$ and the m.e.
$\lan r \ran_{nL}, ~\sqrt{\lan r^2 \ran_{nL}}$~ (in fm) for the potential $V_a(r)$ with $\sigma_a=0.414$~GeV$^2$}
\label{tab.02}
\begin{center}
\begin{tabular}{|c|c|c|c|}
\hline
     State &    $\lan r \ran_{nL}$ (in fm)& $\sqrt{\lan r^2 \ran_{nL}}$ (in fm) & $M_{cog}(nL)$ (in GeV)\\\hline \hline
    $L=0$      &            &       & \\ \hline
     $n_r=0$     & 0.484   & 0.547    &   2.031   \\
     $n_r=1$    & 0.722   & 0.84   &   3.030  \\
     $n_r=2$    & 0.903   & 1.05   &   3.789  \\ \hline
       $L=1$    &            &        &   \\ \hline
     $n_r=0$    &  0.648  &0.70   &  2.718  \\
    $n_r=1$     &  0.837  & 0.95    & 3.511   \\
    $n_r=2$     &  0.994  &1.14    &  4.171   \\ \hline

    $L=2 $     &      &      &     \\ \hline
    $n_r=0$     &  0.779 & 0.83    &  3.268    \\
    $n_r=1$    & 0.940   &1.045    & 3.943  \\
\hline
\end{tabular}
\end{center}
\end{table}
Calculated spin-averaged masses can be compared with those from LQCD calculations \cite{1,2,4}, using their definition for
glueballs: $M_{cog} = \frac{1}{3} M(0^{++}) + 2 M(2^{++})$ (see Table~\ref{tab.03}). In \cite{4} the mass $M_{cog}(n_r=0)=2135$~MeV
is by $\sim 100$~MeV larger than in our work because there a larger string tension, $\sigma_f=0.235$~GeV$^2$, is used.

We pay also attention that the squared masses $M_{cog}^2(n_r=0)$ of the ground states with different $L$ (the solutions of the Eq.~\ref{eq.02})
are described by the leading Regge trajectory for glueballs,
\be
M_{cog}^2(n_r=0) = \sigma_a (8 L  + 3\pi ).~~
\label{eq.10}
\ee
with accuracy better 3\%. It gives the slope of the trajectory equal to  $8\sigma_a =3.312$~GeV$^2$
for $\sigma_a=0.414$~GeV$^2$.

\section{The spin-spin interaction}

The spin-spin potential $V_{ss}$ is usually presented via the function $V_4(r)$, introduced in \cite{25}, and in RSH this potential
\be
V_{ss} = \frac{V_4(r)}{3\mu_n^2}\ves_1\ves_2, ~~
\label{eq.11}
\ee
contains the kinetic energy $\mu_n^2$ in denominator. On fundamental level $V_4(r)$ was studied in \cite{26,27,28},  where its perturbative
part was expressed via the vacuum correlation function $D_1(z)$ (the scalar function):
\be
V_4(r) = \int^\infty_0 {\rm d}\nu ( 6 D_1(z) + 4 r^2 \frac{\partial D_1(z)}{\partial z^2}, ~~(z^2= r^2 + \nu^2).~~
\label{eq.12}
\ee
The correlator $D_1(z)$ was calculated in LQCD at distances $r < 0.20$~fm  and unfortunately, remains undefined in important region,
$r \l 0.20$~fm. Therefore, the potential $V_4(r)$ is also known approximately. The analysis of $D_1(z)$ \cite{28} of the lattice data from \cite{29}
has shown that $V_4(r)$ can be equally good approximated by two expressions:
in case A $V_4(r)$ includes the $\delta$-function (this form is often used in mesons \cite{30}) and in case B
$V_4(r)$ includes  the Yukawa-function as a screening function.

\subsection{Case A}

In adjoint representation,
\be
V_4(r) = 6\alpha_{hf}^{(1)} \delta(r) ~~
\label{eq.13}
\ee
gives the hyperfine splitting,
\be
\delta_{hf}^{(1)}(n,L=0) =  2\alpha_{hf}^{(1)}\frac{|R(0)|^2}{\mu_n^2}, ~~
\label{eq.14}
\ee
where the scale of the coupling $\alpha_{hf}^{(1)}$ is not strictly defined (a fitting parameter). In mesons $\alpha_{hf}$ is usually fitted
by comparison with experimental data, however, in glueballs there is no still consensus, what $f_0$-meson (and $f_2$- meson) is the true scalar
(tensor) glueball \cite{31,32}. Also $\alpha_{hf}^{(1)}$ is not a universal number and may be different in ground and excited states \cite{33},
as well as in different types of mesons. Here in Eq.~\ref{eq.14} we take the coupling at the scale of the vacuum correlation length,
$T_g\sim 0.15(5)$~fm \cite{27,28}, where $\alpha_{\rm V}\cong 0.31(1)$ for $\Lambda_{\overline{MS}}^0=238$~MeV (see Section 2).

In spin-spin splitting the radial w.f. at the origin, $R(0,n) = \frac{\psi(r,n)}{\sqrt{4\pi}}$, and
the gluon  kinetic energy $\mu_n = 1/4 M_{cog}(n)$ were calculated, solving Eq.~\ref{eq.02} ($m_g=0$) with the potential
$V_a(r)=\sigma_a r, ~\sigma_a=0.414$~GeV$^2$. The following $|R(0,n_r)|^2$ were obtained (in GeV$^{3/2}$),
\be
|R (0,n_r=0)|^2 = 0.331; ~~ |R(0,n_r=1)|^2= 0.588;  ~~ |R(0, n_r=2)|^2 =0.756,  ~~
\label{eq.15}
\ee
which grow for higher excitations due to relativistic kinematics. However, the kinetic energy: $\mu_0=0.508$~GeV,
$\mu_1= 0.574$~GeV,~$\mu_2=0.897$~GeV are also growing and due to that $\delta_{hf}^{(1)}(n_r)$ decreases for the higher excitations.
The following values  $\delta_{hf}(n_r) = M(2^{++}) - M(0^{++})$ with  $\alpha_{hf}=0.305$ were calculated,
\be
\delta_{hf}^{(1)}(n_r=0)=784~\text{MeV}, ~~ \delta_{hf}^{(1)}(n_r=1)=625~\text{MeV}, ~~\delta_{hf}^{(1)}(n_r=2) = 513~\text{MeV},~~
\label{eq.16}
\ee

The masses $M(0^{++}) = M_{cog} - \frac{2}{3} \delta_{hf}$ and
$M(2^{++})= M_{cog} + \frac{1}{3} \delta_{hf}$ are given in Table~\ref{tab.03}, together with
those masses from LQCD \cite{1,2,4}. Comparison shows that our number of $\delta_{hf}^{(1)}(n_r=0)=784$~MeV
is close to this splitting $\sim 720(100)$ MeV from \cite{1} and 723~MeV in \cite{4}.

Comparison with the lattice results \cite{4} shows that in case A the mass difference - 625~MeV for first excitation is
significantly larger than $\delta_{hf}(n_r=1)=460(90)$~MeV in \cite{4}, although there absolute values of $M(2^{++}), M(2^{*++})$ \cite{4}
are by $\sim 100$~MeV larger, due to a larger string tension $\sigma_a~(\sigma_f)$  used.

In recent reviews \cite{31,32}  $f_0(1500)$ and $f_2(2340)$ mesons are supposed to be main candidates for scalar and
tensor glueballs (the ground states) and for them  experimental value, $\delta_{hf}(exp.)=M(f_2(2330)) - M(f_0(1500))=830(50)$~MeV,
(the masses are taken from PDG \cite{34}) is even a bit larger than  our number 784~MeV. It is of interest that in our work
$M(0^{++})=1508$~MeV coincides with the mass of $f_0(1500)$ meson ($M(f_0(1500))= 1522(25)$~MeV \cite{34}). Also  in case A
$M(2^{++})=2292$~MeV coincides with the mass of $f_2(2300)$ meson,  $M(f_2(2300))=2297(28)$~MeV, and also is close
to the mass of very wide resonance $f_2(2340)$ with the width $\Gamma\sim 330$~MeV.

In theoretical models very different  $\delta_{hf}$ were predicted:
887 MeV \cite{7}, 730(310) MeV \cite{35},  440 MeV\cite{10} (see Table~\ref{tab.04}).

\subsection{Case B}

In case B the potential $V_4(r)$ contains a screening factor $U(r)$ \cite{25,26}:
\be
V_4(r) = 6 \alpha_{hf}^{(2)} U(r), ~~
\label{eq.17}
\ee
with  $U(r)= C_s\frac{\exp(-\beta r)}{r}$, normalized as
\be
\int^\infty_0 {\rm d}r r^2 U(r) =1.0.
\label{eq.18}
\ee
It gives $C_s=\beta^2$ and here $\beta=2.463$~GeV is taken  from \cite{28}, where the lattice data \cite{29} were parameterized.
In the potential Eq.~\ref{eq.17} $\beta$ defines the scale of $\alpha_{hf}^{(2)}$:
$r_{s2}=\beta^{-1}=0.08$~fm, which is small and  $\alpha_{hf}^{(2)}(r_{s2})=\alpha_{\rm V}=0.255$ for $\Lambda_{\overline{MS}}^0=238$~MeV
and $M_B=1.0$~GeV (see Section 2). Note that this scale is smaller than that in lattice data \cite{36}, where it is equal to the vacuum correlation length
$T_g\sim 0.22$~fm and the coupling is larger, $\alpha_{\rm V}(T_g)\sim 0.33$.

The matrix elements over $V_4(r)$  Eq.~\ref{eq.17},
\be
V_n=\lan V_4(r) \ran_n = 6 \alpha_{hf}^{(2)}C_s\int^\infty_0 {\rm d}r r |R(r,n)|^2 \exp(-\beta r),   ~~
\label{eq.19}
\ee
have the following values (in GeV$^3$):
\be
\lan V_4 \ran_n = 1.641~(n_r=0), ~~2.587~ (n_r=1),~2.908~(n_r=2).
\label{eq.20}
\ee
Then with $\alpha_{hf}^{(2)}=0.255$  the hyperfine splitting,
\be
\delta_{hf}^{(2)}(n_r) = \frac{V_n}{3\mu_n^2},~~
\label{eq.21}
\ee
 have the values:
\be
\delta_{hf}(n_r=0)=543~\text{MeV};~~ \delta_{hf}(n_r=1)= 383~\text{MeV};~~ \delta_{hf}(n_r=2)= 276~\text{MeV},  ~~
\label{eq.22}
\ee
which are  $\sim (30-40)\%$ smaller than in case A. Note that in case B hyperfine splittings decrease mostly because of the screening factor, present in the correlation function $D_1(z)$.

Calculated $\delta_{hf}^{(2)}=543$~MeV occurs to be close to the value 440~MeV in \cite{10} and 451~MeV in \cite{37}.
Thus in case B the ground state mass $M(2^{++})=2212$~MeV
is by 80~MeV smaller than in case A and coincides with the mass of $f_2(2210)$ meson.
Note that from the analysis of $J/\psi$ radiative decays the meson $f_2(2210)$ is supposed to be  a tensor glueball
\cite{37}.

\begin{table}
\caption{The masses of scalar and tensor glueballs (in MeV) in RSH with linear potential ($\sigma_a=0.414$~GeV$^2$, $N_f=0$)}
\label{tab.03}
\begin{center}
\begin{tabular}{|c|c|c|c|c|c|}
\hline
\multirow{3}{*}{States}& \multirow{3}{*}{\cite{1}}&\multirow{3}{*}{\cite{2}}&\multirow{3}{*}{\cite{4}}&Case A: & Case B: \\
& & & & $\alpha_{hf}^{(1)}=0.305$ & $\alpha_{hf}^{(2)}=0.255$\\
& & & & & in Eq.~(\ref{eq.21})\\
\hline \hline
   $\sigma_f$ (in GeV$^2$)  & 0.1936  & 0.1936      &   0.235    &  0.184             & 0.184   \\

 $0^{++},n_r=0$           &1550(100)   & 1730(50)   & 1653(26)     & 1508             & 1669 \\

 $0^{++},n_r=1$         &            &             & 2842(40)      &   2613  &         2775   \\

 $0^{++},n_r=2$         &           &              & 3650(60)     &   3447   &     3605         \\

  $2^{++},n_r=0$         & 2270(100) & 2400(25)     &   2376(32)     & 2292          & 2212   \\

 $2^{++},n_r=1$         &            &             &   3300(50)      &  3238     & 3158  \\

 $2^{++},n_r=2$           &            &            &         & 3960       & 3881 \\
 \hline
 \end{tabular}
 \end{center}
 \end{table}
In theoretical models (see Table~\ref{tab.04}) the masses of scalar and tensor glueballs vary in large interval,
$\sim (200-400)$~MeV both for the ground and excited states. Such large discrepancies demonstrate that the glueball masses are very sensitive
to the model and parameters used. For example, in \cite{38} the tensor glueball mass of the ground state, equal to $\sim 3100$~MeV, is predicted
within the sum rule approach.

\begin{table}
\caption{The masses of the scalar and tensor glueballs (in GeV)  in different theoretical models}
\label{tab.04}
\begin{center}
\begin{tabular}{|c|c|c|c|c|c|}
\hline
\multirow{2}{*}{States}& \multirow{2}{*}{\cite{10}}&\multirow{2}{*}{\cite{7}}&\multirow{2}{*}{\cite{35}}& Case A: & Case B: \\
& & & & [this work] & [this work] \\
\hline \hline
$0^{++}$ &     1.98      &  1.710     & 1.800 (120)  &1508   & 1.669 \\
$0^{*++}$  &    3.36     &  2.631     &  2.500(210)  & 2613    & 2.775  \\
$0^{**++}$  &             &          & 3.610(150)  &  3447   &  3.605  \\
$2^{++}$ &    2.42    &    2.597    &    2.530(180)  & 2292   & 2.212  \\
$2^{*++}$  &  3.11   &   2.912      &   3.530(230)  &  3238   &  3.158  \\
$2^{**++}$  &        &              &               & 3960   &  3.881  \\
\hline
\end{tabular}
\end{center}
\end{table}

\section{Conclusions}

We studied the scalar and tensor glueballs in pure gauge theory and first of all, calculated spin-averaged masses of the ground
and excited states, which are very useful  for the analysis, because they depend only on the single parameter -- the string
tension $\sigma_a=0.414$~GeV$^2$ in adjoint representation. The value of $\sigma_a$ taken corresponds to $\sigma_f=0.184$~GeV$^2$
in FR and was extracted from the lattice data. In the spin-spin potential $V_{ss}$ additional parameter --
the strong coupling $\alpha_{hf}$ is present, which is not fixed yet. Moreover, the functional form of $V_{ss}$ is not known in strict sense:
it can be very short-ranged potential and includes $\delta$-function (case A), or has more soft behavior as the
Yukawa-function.

In case A  the scale of $\alpha_{hf}^{(1)}$ is defined by the only one parameter -- $\Lambda_{\overline{MS}}^0= 238$~MeV~($N_f=0$) \cite{20}
and for the ground states $\delta_{hf}^{(1)}=784$~MeV, $M(0^{++})=1508$~MeV, and $M(2^{++})=2292$~MeV are obtained.
It is of interest that these masses coincide with experimental masses of $f_0(1500)$ and $f_2(2300)$  mesons,
or the mass of $f_2(2340)$ meson, which width  $\Gamma\sim 330$~MeV is very large. Note that in \cite{31,32}
$f_0(1500),~f_2(2340)$ are considered as main glueball candidates.

In case B, due to screening factor, the hyperfine splitting $\delta_{hf}=543$~MeV is by 30\% smaller
and it gives  $M(0^{++})=1669$~MeV,  $M(2^{++})=2212$~MeV. Then with this type of the spin-spin dynamics
the glueball ground states  may be associated with $f_0(1710)$ and $f_2(2210)$. To make choice between
two variants additional analysis of experimental data is needed.

For first excitation our calculations give $M(0^{*++})=2613$~MeV~(in case A) and 2775~MeV~(in case B) and
for second excitation $M(0^{**++})\cong 3.90$~GeV is predicted.



\end{document}